\documentclass[twocolumn,pre,superscriptaddress]{revtex4}

\usepackage{graphicx}
\usepackage{amsmath,amssymb}
\usepackage{color}
\usepackage{calrsfs}

\newcommand\defn{\textit}
\newcommand\mat{\mathbf}

\newcommand\dd{\mathrm{d}}
\newcommand\ii{\mathrm{i}}
\renewcommand\vec{\mathbf}
\newcommand\av[1]{\langle#1\rangle}
\newcommand\etal{\textit{et~al.}}
\newcommand\half{\tfrac12}
\newcommand{\Ord}{\mathrm{O}}
\renewcommand{\Im}{\operatorname{Im}}

\newcommand{\tr}{\operatorname{Tr}}

\begin{document}

\title{Spectra of random graphs with arbitrary expected degrees}

\author{Raj Rao Nadakuditi}
\affiliation{Department of Electrical Engineering and Computer Science,
  University of Michigan, Ann Arbor, MI 48109}
\author{M. E. J. Newman}
\affiliation{Department of Physics and Center for the Study of Complex
  Systems, University of Michigan, Ann Arbor, MI 48109}

\begin{abstract}
  We study random graphs with arbitrary distributions of expected degree
  and derive expressions for the spectra of their adjacency and modularity
  matrices.  We give a complete prescription for calculating the spectra
  that is exact in the limit of large network size and large vertex
  degrees.  We also study the effect on the spectra of hubs in the network,
  vertices of unusually high degree, and show that these produce isolated
  eigenvalues outside the main spectral band, akin to impurity states in
  condensed matter systems, with accompanying eigenvectors that are
  strongly localized around the hubs.  We also give numerical results that
  confirm our analytic expressions.
\end{abstract}


\maketitle

\section{Introduction}
\label{sec:intro}
The topology of complex networks, such as social, biological, and
technological networks, can be represented in matrix form using an
adjacency matrix or any of several other related matrices such as the graph
Laplacian or the modularity matrix~\cite{Newman03d,Boccaletti06}.  The
spectral properties of these matrices---their eigenvalues and
eigenvectors---are related to a range of network features of scientific
interest, including optimal partitions~\cite{Fiedler73,PSL90}, percolation
properties~\cite{BBCR10}, community structure~\cite{Newman06b,Fortunato10},
and the behavior of network dynamical processes such as random walks,
current flow, diffusion, and synchronization~\cite{Arenas08,BBV08}.  As a
result, the study of network spectra has been the subject of considerable
research effort for some years.  This effort has taken a number of forms.
One has been the study of the spectra of empirically observed networks,
which can be calculated by numerical means for networks of size up to
hundreds of thousands of vertices~\cite{Arenas08,FDBV01}.  Another, which
is the topic of this paper, is the study of the spectra of model networks.
A fundamental question we would like to answer is how particular structural
features of networks are reflected in network spectra, and model networks
provide an ideal setting in which to investigate this question.

Some results in this area have been known for a long time.  For example,
the very simplest of network models, the Poisson random graph, studied as
far back as the 1950s by Erd\H{o}s, R\'enyi, and others~\cite{SR51,ER60},
has a symmetric adjacency matrix whose elements are independent
identically-distributed random variables.  Such matrices are known, subject
to some conditions but regardless of the precise distribution of their
elements, to have a universal spectrum obeying the Wigner semicircle law,
and eigenvectors that are distributed isotropically at random, except for
the leading eigenvalue and eigenvector, whose values are governed by the
Perron--Frobenius theorem and the average degree of the
network~\cite{arnold1971wigner,furedi1981eigenvalues,anderson2010introduction,bai2010spectral,vanMieghem11,erdos2011spectral,tao2011topics,tao2012random}.

As we have come to understand in the last decade, however, the random graph
is a poor model for the structure of real-world networks.  In particular,
the frequency distribution of the degrees of vertices in the random graph
is Poissonian, while the degree distribution of most real-world networks is
highly right-skewed, often having a power-law or exponential tail of
``hubs'' with degree far above the mean.  Luckily, it turns out to be
possible to create generalizations of the basic random graph that
incorporate arbitrary degree distributions, including skewed distributions,
the best-known such model being the so-called configuration
model~\cite{Bollobas80,MR95}.  The configuration model is solvable exactly
for many of its structural properties, including its complete component
structure~\cite{MR95,MR98,NSW01} and percolation
properties~\cite{CEBH00,CNSW00}, and the results have led us to a better
understanding of the profound effect the degree distribution has on network
topology.

In this paper we study the spectral properties of the configuration model.
Motivated by recent developments in random matrix theory, we derive a
simple recipe for calculating the spectrum of the adjacency matrix of the
model.  We show that the spectrum is composed of three fundamental
elements, all of which have clear correlates in the structure of the
network.  The elements are: (1)~the leading eigenvalue, which is dictated
primarily by the average network degree; (2)~a continuous band or ``bulk
spectrum,'' analogous to the Wigner semicircle but taking a different
shape; and (3)~in some but not all cases, additional eigenvalues outside of
the continuous band which correspond to the hubs in the network and which
have eigenvectors that are strongly localized about those hubs.

In addition to our analytic developments, we also confirm the form and
behavior of each of these elements with numerical calculations on example
networks generated using the configuration model.

A number of previous authors have examined the spectral properties of the
configuration model.  Farkas~\etal~\cite{FDBV01} performed numerical
calculations on large samples generated using the model and demonstrated
that there are clear deviations from the semicircle law for non-Poisson
choices of the degree distribution, and especially for power-law
distributions.  Dorogov\-tsev~\etal~\cite{DGMS03} gave an analytic route to
the full spectrum, though their method is complex, involving the solution
of a nonlinear integral equation containing Bessel functions, which at
present can only be done approximately.  Chung~\etal~\cite{CLV03} gave a
rigorous derivation of the expected value of the largest eigenvalue in the
spectrum in the limit of a dense network.  Our calculations extend these
studies by providing a simple derivation of the full spectrum which is
exact in the limit of large vertex degrees and confirms earlier findings
while shedding new light on features of the spectrum and their implications
for network structure.

\section{The model}
In this paper we study the spectral properties of the configuration
model---or, more precisely, a slight variant of the model, as we now
describe.

The configuration model is a model of an undirected random graph with a
specified number of vertices~$n$ and a given degree sequence.  In this
model one first specifies a degree sequence, meaning one specifies the
degree of each of the $n$ vertices.  Let the degree of vertex~$i$ be
denoted~$k_i$ and let us visualize the degree as $k_i$ ends or ``stubs'' of
edges emerging from the vertex.  Then the configuration model is defined as
the ensemble of pairwise matchings of stubs in which every matching appears
with equal probability.  That is, a configuration model network with the
given degree sequence is generated by repeatedly choosing two stubs
uniformly at random from those available and joining them together to form
a complete edge.  This process continues until all stubs have been joined
and no unattached stubs remain.  (For this to work, the number of stubs
must be even, and hence the model is defined only for degree sequences
whose sum $\sum_i k_i$ is even.)

The configuration model provides a way to generate networks that have any
degree sequence we desire while being essentially random in other
respects---there are no correlations or long-range structure in the
configuration model ensemble.

A crucial feature of the configuration model for our purposes will be the
expected number of edges between a vertex pair.  It is straightforward to
show, given the degree sequence, that the expected number of edges between
vertices~$i$ and~$j$ is equal to $k_ik_j/2m$ in the limit of large network
size, where $m=\half \sum_i k_i$ is the number of edges in the network.
Note that it is possible to generate networks with multi-edges---pairs of
vertices connected by more than one parallel edge.  The actual number of
edges between vertices~$i$ and~$j$ is multinomially distributed with mean
$k_ik_j/2m$.

However, edges in the configuration model are not statistically
independent.  Since the degrees of vertices are fixed, the presence of an
edge from vertex~$i$ to vertex~$j$ makes it less likely that there will be
an edge from~$i$ to any other vertex, and hence edges that share a common
end are correlated.  When degree is large the correlations become small and
the multinomial distribution of edge number becomes approximately Poisson,
but for networks with finite average degree the correlations will always be
present and may be significant.

These correlations make analysis of the model more difficult and so in this
paper we consider a modified model in which the number of edges between
each pair of vertices is \emph{defined} to be an independent random
variable with mean $k_ik_j/2m$ and value drawn from a Poisson distribution
with that mean.  In this model, $k_i$~becomes the expected degree of
vertex~$i$ and $m$ is the expected total number of edges.  When degrees
become large, which is the primary regime that we consider in this paper,
the actual degrees will be narrowly peaked about their expected values, so
the properties of the variant model and the standard configuration model,
including the spectral properties that we study, become the same.  This
model (or slight variants of it) has been studied previously by a number of
authors, notably Chung and Lu~\cite{CL02b}, with whose work it is perhaps
most strongly associated.

In this paper we consider networks in the limit of large size~$n$ with
expected vertex degrees drawn from a fixed probability density~$p(k)$, so
that $p(k)\>\dd k$ is the fraction of vertices with expected degree in the
interval from $k$ to $k+\dd k$.  (Note that expected degree need not be an
integer, although one is free to choose it to have integer values if one
wishes.)  More precisely, we consider a sequence of networks of increasing
size with fixed expected degrees~$k_i$ and additional degrees drawn
from~$p(k)$ as $n$ becomes larger.  Thus for finite $n$ the expected degree
of any particular vertex~$i$ remains constant as $n$ becomes large and the
empirical degree distribution converges to $p(k)$ in the large-$n$ limit.

The adjacency matrix~$\mat{A}$ of a network generated according to this
model is the $n\times n$ symmetric matrix with integer elements $A_{ij}$
equal to the number of edges between vertices~$i$ and~$j$.  Our primary
goal in this paper is to calculate the average spectrum of the adjacency
matrix within the model ensemble, which we do in two stages.  We write the
matrix as
\begin{equation}
\mat{A} = \av{\mat{A}} + \mat{B},
\end{equation}
where $\av{\mat{A}}$ is the ensemble average of~$\mat{A}$, which has
elements $\av{A_{ij}} = k_ik_j/2m$, and $\mat{B}$ is the deviation from
that average.  Our approach is first to calculate the spectrum of the
matrix~$\mat{B}$, whose elements are, by definition, independent random
variables with zero mean, although crucially they are are not identically
distributed.  Once we have the spectrum of~$\mat{B}$ then the spectrum of
$\mat{A}$ is calculated from it in a separate step.

The matrix~$\mat{B}$ is of interest in its own right.  It has elements
\begin{equation}
B_{ij} = A_{ij} - \av{A_{ij}} = A_{ij} - {k_ik_j\over2m}.
\label{eq:mm}
\end{equation}
This matrix is known as the modularity matrix, and forms the basis for one
of the most widely used methods for detecting modules or communities in
networks~\cite{Newman06b,Fortunato10}.  The methods described in this paper
thus give us the spectra of both the adjacency matrix and the modularity
matrix.

Note that the elements of the modularity matrix have variance the same as
the elements of the adjacency matrix which, since they are Poisson
distributed, have variance equal to their mean $k_ik_j/2m$.  Hence
\begin{equation}
\bigl\langle B_{ij}^2 \bigr\rangle = {k_ik_j\over2m},
\label{eq:bvar}
\end{equation}
which will be important shortly.

\section{Spectrum of the modularity matrix}
\label{sec:spectrum}
As discussed in the previous section, we will first calculate the spectrum
of the modularity matrix~$\mat{B}$, defined by Eq.~\eqref{eq:mm}, then
calculate the spectrum of the adjacency matrix from it in a separate step.
We begin by developing some fundamental notions concerning random variables
that will be important for our derivations.

Suppose we have two independent random variables, $x$ and~$y$, ordinary
scalar variables, with probability densities~$p_x(x)$ and~$p_y(y)$.  What
is the probability that their sum~$x+y$ will have a particular value~$z$?
The answer to this question is well known and simple.  The probability
density for~$z$ is
\begin{align}
p(z) &= \iint p_x(x) p_y(y) \delta(x+y-z) \>\dd x\,\dd y \nonumber\\
     &= \int p_x(x) p_y(z-x) \>\dd x
\label{eq:addconv}
\end{align}
which is the convolution of the two distributions.  Similarly we can ask
for the probability that the product~$xy$ has value~$z$, which is given by
the multiplicative convolution
\begin{align}
p(z) &= \iint p_x(x) p_y(y) \delta(xy-z) \>\dd x\,\dd y \nonumber\\
     &= \int p_x(x) p_y(z/x) \>{\dd x\over x}.
\label{eq:multconv}
\end{align}

A scalar random variable can be thought of as the single eigenvalue of a
$1\times1$ random matrix.  A~$1\times1$ matrix is diagonal by definition
and its one eigenvalue is trivially equal to its one element.  A natural
generalization of the convolution results above is to ask what their
equivalent is for larger random matrices, $2\times2$, $3\times3$, and so
forth, where we will confine ourselves to symmetric matrices, so that the
eigenvalues are real.  That is, if we know the probability density of the
eigenvalues---the so-called spectral density---of two independent symmetric
random matrices, what is the spectral density of their sum or product?  The
answer is no longer a simple convolution, because matrices do not in
general commute, so what is the appropriate generalization?  Unfortunately,
this question does not have a straightforward answer because it turns out
that a knowledge of the spectral densities alone is not enough.  In general
one needs to know the distribution of the entire matrices to calculate the
spectral density of their sum or product.  There is, however, one case in
which relatively simple results apply, which is when the eigenvectors of
the two matrices are themselves random and uncorrelated.

Recall that the eigenvectors of a symmetric matrix are orthogonal---for an
$n\times n$ matrix they define a set of orthogonal axes in an
$n$-dimensional vector space.  Thus if we have two random symmetric
matrices, the eigenvectors of one can always be transformed into the
eigenvectors of the other by a suitable rotation and/or reflection---in
other words by a suitable unitary transformation.  If for different choices
of the random matrices the transformations needed to do this are
distributed isotropically---if all possible such transformations are
equally likely---then the random matrices are said to be \defn{free}.
Loosely, one can say that two random matrices are free if the angle between
their eigenvectors is also random.  The mathematics of free random
variables has been developed extensively since the 1990s and is known by
the name of \defn{free probability theory}~\cite{voiculescu1992free}.

The crucial observation now is the following: for free matrices the
spectral density of their sum or product is a function only of the
individual spectral densities.  It turns out that one no longer needs to
know the entire distribution of the matrices themselves and well-defined
generalizations of the convolution equations, Eqs.~\eqref{eq:addconv}
and~\eqref{eq:multconv}, exist.  For the sum of two matrices the
appropriate generalization is known as the \defn{free convolution} or
\defn{free additive convolution}; for the product of matrices it is the
\defn{free multiplicative convolution}.  Thus if two symmetric random
matrices have spectral densities~$p_x(x)$ and $p_y(y)$, then the spectral
density of their product is the free multiplicative convolution
\begin{equation}
p(z) = (p_x\boxtimes p_y)(z),
\label{eq:freemult}
\end{equation}
where $\boxtimes$ denotes the convolution.  Although this defines the
convolution in principle, it does not tell us how to calculate it.  We will
come to that in a moment, but first let us return to the configuration
model and see why this is a useful result.

We wish to calculate the spectral density of the modularity
matrix~$\mat{B}$, which for an undirected network is a symmetric random
matrix whose elements have zero mean but different variances, equal
to~$k_ik_j/2m$---see Eq.~\eqref{eq:bvar}.  Let us define a normalized
modularity matrix~$\widetilde{\mat{B}}$ by
\begin{equation}
\widetilde{\mat{B}} = \mat{D}^{-1/2} \mat{B} \mat{D}^{-1/2},
\label{eq:equivalentB}
\end{equation}
where $\mat{D}$ is the diagonal matrix with elements~$k_i$.
$\widetilde{\mat{B}}$~has elements~$\widetilde{B}_{ij} =
B_{ij}/\sqrt{k_ik_j}$, so that each is divided by a factor proportional to
its standard deviation and hence, though not identically distributed, all
elements now have the same variance, equal to~$1/2m$.  So long as the
vertex degrees are large, matrices with this property are known to have an
eigenvector basis oriented isotropically at random and to have spectral
density obeying the Wigner semicircle
law~\cite{arnold1971wigner,furedi1981eigenvalues,anderson2010introduction,bai2010spectral,vanMieghem11,erdos2011spectral,tao2011topics,tao2012random},
which for our particular matrix takes the form
\begin{equation}
\rho_c(z) = {1\over2\pi} \sqrt{4c-c^2z^2},
\label{eq:rho1}
\end{equation}
where $c=2m/n$ is the average degree in the network.  The requirement that
vertex degrees be large is necessary because deviations from the semicircle
law are known to arise for very sparse matrices~\cite{RB88}.  For small
degrees, therefore, the results given here will only be approximate.

Now consider an eigenvalue~$z$ of the modularity matrix~$\mat{B}$ itself,
satisfying $\mat{B}\vec{b} = z\vec{b}$ where $\vec{b}$ is the corresponding
eigenvector.  Multiplying by $\mat{D}^{1/2}$, writing
$\mat{B}=\mat{D}^{1/2} \widetilde{\mat{B}} \mat{D}^{1/2}$, and defining
$\vec{v} = \mat{D}^{1/2}\vec{b}$, this can also be written
\begin{equation}
\mat{D}\widetilde{\mat{B}}\vec{v} = z\vec{v}.
\end{equation}
In other words the modularity matrix has the same eigenvalues as the matrix
$\mat{D}\widetilde{\mat{B}}$, which is the product of the diagonal
matrix~$\mat{D}$, which by definition has spectral density equal to the
degree distribution~$p(k)$, and the symmetric matrix~$\widetilde{\mat{B}}$,
with spectral density~$\rho_c(z)$ given by Eq.~\eqref{eq:rho1}.

But it is precisely to the products of such random matrices that
Eq.~\eqref{eq:freemult} relates, and hence, applying that equation, we
arrive at the principal result of this paper: the spectral density of the
modularity matrix for a network with arbitrary expected degrees is equal to
the free multiplicative convolution of the degree distribution with the
Wigner semicircle.  That is, the spectral density~$\rho(z)$ is given by
\begin{equation}
\rho(z) = (p \boxtimes \rho_c)(z),
\label{eq:principal}
\end{equation}
where $p(k)$ is the distribution of expected degrees and $\rho_c(z)$ is
given by Eq.~\eqref{eq:rho1}.

This result is of immediate practical utility.  Numerical methods exist for
computing free multiplicative convolutions
efficiently~\cite{rao2008polynomial,olver2012numerical}, which means we can
use existing numerical packages to compute spectral densities easily and
rapidly for a wide range of degree distributions.

For the purposes of the present paper, however, we would like to know more.
In particular, we would like explicit formulas for calculating the spectral
density in the general case.  Unfortunately, the free multiplicative
convolution has no simple expression for matrices of finite size, but in
the limit of large size---which is also the limit of a large
network---suitable expressions do exist.  Specifically, for a spectral
density~$\rho$ we can define a function
\begin{equation}
\Gamma_\rho(z) = \int {x\,\rho(x)\>\dd x\over z-x},
\label{eq:defgamma}
\end{equation}
which is called the \defn{Cauchy transform} of~$x\rho(x)$.  Then if $\rho$
is the free multiplicative convolution of two other distributions $p$ and
$\rho_c$ as in Eq.~\eqref{eq:principal}, it can be shown that
\begin{equation}
\Gamma_\rho^{-1}(u) = {u\over u+1} \Gamma_p^{-1}(u)
                      \Gamma_{\rho_c}^{-1}(u),
\label{eq:gammagamma}
\end{equation}
where $\Gamma^{-1}$ denotes the functional inverse of~$\Gamma$, and
$\Gamma_p$ and $\Gamma_{\rho_c}$ are defined by analogy
with~\eqref{eq:defgamma}:
\begin{equation}
\Gamma_p(z) = \int {k\,p(k)\>\dd k\over z-k},\quad
\Gamma_{\rho_c}(z) = \int {x\,\rho_c(x)\>\dd x\over z-x}.
\label{eq:othergamma}
\end{equation}

Substituting Eq.~\eqref{eq:rho1} into the second of these, we have
\begin{align}
\Gamma_{\rho_c}(z) &= {1\over2\pi} \int_{-2/\sqrt{c}}^{2/\sqrt{c}}
                      {x\sqrt{4c-c^2x^2}\over z-x} \>\dd x \nonumber\\
  &= \half cz \bigl(z \pm \sqrt{z^2-4/c} \bigr) - 1.
\label{eq:deriv1}
\end{align}
The ambiguity in the sign of the square root arises because of a branch cut
in the evaluation of the integral, but it can be shown that the final
result for the free convolution never depends on the choice of
sign~\cite{rao2007multiplication}.  Here we take the negative sign, since
it makes some of the following steps cleaner.  Rearranging for $z$ as a
function of $\Gamma_{\rho_c}$ we then find that the functional inverse is
\begin{equation}
\Gamma_{\rho_c}^{-1}(u) = {u+1\over\sqrt{cu}},
\end{equation}
and substituting into~\eqref{eq:gammagamma} we get
\begin{equation}
\Gamma_\rho^{-1}(u) = \sqrt{u\over c}\,\Gamma_p^{-1}(u).
\end{equation}
Evaluating this equation at the point $u=\Gamma_\rho(z)$ gives $z =
\sqrt{\Gamma_\rho(z)/c} \>\Gamma_p^{-1}(\Gamma_\rho(z))$, which can be
rearranged to read
\begin{equation}
\Gamma_\rho(z) = \Gamma_p\Bigl( z \sqrt{c/\Gamma_\rho(z)} \Bigr).
\label{eq:gammarho}
\end{equation}
For convenience we define $h(z) = \sqrt{\Gamma_\rho(z)/c}$ and
Eq.~\eqref{eq:gammarho} becomes
\begin{equation}
c h^2(z) = \Gamma_p\bigl( z/h(z) \bigr)
         = \int_0^\infty {k\,p(k)\>\dd k\over z/h(z) - k},
\end{equation}
or, more simply,
\begin{equation}
h(z) = {1\over c} \int_0^\infty {k\,p(k)\>\dd k\over z - k h(z)}.
\label{eq:fptsolution1}
\end{equation}

If we can solve this equation for~$h(z)$ then the Cauchy transform
of~$x\rho(x)$, Eq.~\eqref{eq:defgamma}, is given by~$\Gamma_\rho(z) =
ch^2(z)$.  To recover $\rho$ itself from the Cauchy transform we note that
for real $x$ and~$\eta$
\begin{equation}
-{1\over\pi} \Im {1\over x+\ii\eta} = {\eta/\pi\over x^2+\eta^2},
\end{equation}
which is a Lorentzian of width~$\eta$ and area~1, and hence in the limit as
$\eta\to0^+$ becomes equal to a delta-function:
\begin{equation}
-{1\over\pi} \lim_{\eta\to0+} \Im {1\over x+\ii\eta} = \delta(x).
\label{eq:delta}
\end{equation}
Thus
\begin{align}
z \rho(z) &= \int x \rho(x) \delta(z-x) \>\dd x \nonumber\\
  &= -{1\over\pi} \lim_{\eta\to0+} \Im \int {x \rho(x)\over z - x + \ii\eta}
      \>\dd x \nonumber\\
  &= -{1\over\pi} \lim_{\eta\to0+} \Im \Gamma_\rho(z+\ii\eta).
\end{align}
This is the \defn{Stieltjes--Perron inversion formula}.  Setting
$\Gamma_\rho(z) = ch^2(z)$ it tells us that the spectral density of the
configuration model is given by
\begin{equation}
\rho(z) = - {c\over\pi z} \Im h^2(z),
\label{eq:fptsolution2}
\end{equation}
where the imaginary part is taken in the limit as~$z$ tends to the real
line from above.

Equations~\eqref{eq:fptsolution1} and~\eqref{eq:fptsolution2} give us a
complete recipe for calculating the spectrum of the modularity matrix.  We
note that equations equivalent to these have been derived in other contexts
in the literature on random matrices.  See for example the results on band
matrices in
Refs.~\cite{molchanov1992limiting,shlyakhtenko1996random,anderson2006clt,casati2009wigner,bai2010limiting}.

\subsection{Example solutions}
\label{sec:solution}
The solution of Eqs.~\eqref{eq:fptsolution1} and~\eqref{eq:fptsolution2}
relies on our being able to compute the integral in
Eq.~\eqref{eq:fptsolution1}, whose difficulty depends on the particular
choice of degree distribution.  To give an example where the calculation is
straightforward, consider the standard Poisson random graph, for which all
vertices have the same expected degree~$c$ and hence $p(k) = \delta(k-c)$.
Substituting into~\eqref{eq:fptsolution1} and solving the resulting
quadratic equation gives
\begin{equation}
h(z) = {z - \sqrt{z^2-4c}\over 2c},
\label{eq:poissonhz}
\end{equation}
so that the spectral density is
\begin{equation}
\rho(z) = {\sqrt{4c-z^2}\over2\pi c},
\end{equation}
which recovers the standard semicircle distribution for the random graph.

As a more general example, consider any distribution where the degrees take
a set of~$\ell$ discrete values~$d_r$, as they do for any integer-valued
degree distribution of the type commonly considered for network models.
Then $p(k) = \sum_{r=1}^\ell p_r \delta(k-d_r)$, where the
coefficients~$p_r$ satisfy~$\sum_r p_r = 1$.  Then, from
Eq.~\eqref{eq:fptsolution1},
\begin{equation}
h(z) = {\sum_{r=1}^\ell p_r d_r/[z - d_r h(z)]\over
        \sum_{r=1}^\ell p_r d_r}
\end{equation}
where we have used $c = \sum_r p_r d_r$.  Thus $h(z)$ is the root of a
polynomial of degree~$\ell+1$.  For instance, if there are two discrete
values of the expected degree, then
\begin{equation}
h(z) = {1\over p_1d_1+p_2d_2} \biggl[ {p_1d_1\over z-d_1 h(z)}
       + {p_2d_2\over z-d_2 h(z)} \biggr],
\label{eq:twodegree}
\end{equation}
which can be rearranged to give the cubic equation
\begin{equation}
 d_1d_2 h^3 - (d_1+d_2) zh^2
 + \biggl[ {d_1d_2\over p_1d_1+p_2d_2} + z^2 \biggr]h - z = 0.
\label{eq:cubic}
\end{equation}
Of the three solutions to this equation one is always real, and hence (in
light of Eq.~\eqref{eq:fptsolution2}) cannot give the spectral density.
The remaining two are complex conjugates and so give results that differ
only in sign, the positive sign being the one we are looking for.

\begin{figure}
\begin{center}
\includegraphics[width=\columnwidth]{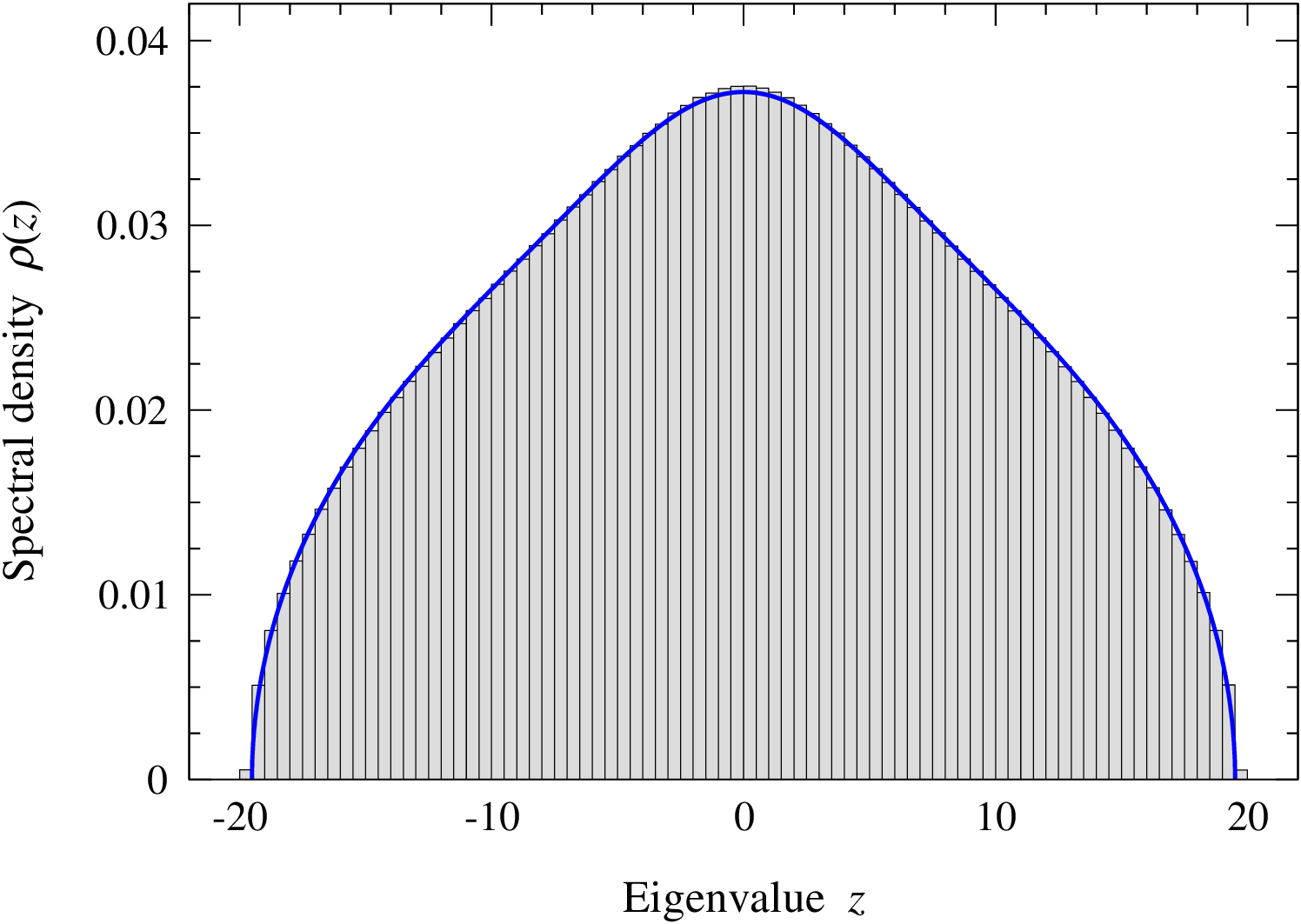}
\end{center}
\caption{The spectral density~$\rho(z)$ for the degree distribution
  described in the text, in which vertices have expected degree $d_1=50$
  with probability $p_1=\frac14$ and $d_2=100$ with probability
  $p_2=\frac34$.  The curve gives the analytic solution, derived from
  Eqs.~\eqref{eq:fptsolution2} and~\eqref{eq:cubic}; the histogram shows
  the results of numerical calculations for actual networks with
  $n=10\,000$ vertices, averaged over $100$ different networks.}
\label{fig:d2spect}
\end{figure}

Figure~\ref{fig:d2spect} shows a plot of the resulting spectral density,
Eq.~\eqref{eq:fptsolution2}, as a function of~$z$ for the case $d_1=50$,
$d_2=100$, and $p_1=1-p_2=\frac14$.  The figure shows strong departure from
the semicircle law.  Also shown are the results of direct numerical
calculations of the spectra of simulated networks with the same degree
distribution and the agreement between the analytic and numerical results
is good.

\subsection{Features of the spectral density}
\label{sec:features}
We can invoke additional properties of the free convolution to better
understand the spectrum of the modularity matrix.  Consider, for instance,
the case where the expected degree distribution~$p(k)$ has compact support,
meaning that there are hard upper and lower limits to the expected degree a
vertex may have.  (The lower limit is trivial, since degrees must be
non-negative, but the upper limit is not.)  Since the semicircle
distribution, Eq.~\eqref{eq:rho1}, also has compact support, the spectral
density of the modularity matrix is then a convolution of two compact
distributions.  In this scenario it can be shown that the bulk spectrum of
the modularity matrix will also have compact
support~\cite{bai2010limiting}.  Furthermore, given this observation we can
show that the spectrum will generically exhibit a sharp square-root decay
at its edges.  To see this, note that the central function~$h(z)$ in our
theory is the solution for $h$ of an equation of the form~$f(h,z)=0$ where
$z$ is given and
\begin{equation}
f(h,z) = {1\over c} \int_0^\infty {k\,p(k)\>\dd k\over z - k h} - h.
\end{equation}
(See Eq.~\eqref{eq:fptsolution1}.)  From Eq.~\eqref{eq:fptsolution2} we
know that $h$ is complex within the spectral band and real outside it and
hence the edge of the band is the point at which complex solutions to
$f(h,z)=0$ disappear.  For analytic~$f(h,z)$ such a disappearance
corresponds to the point at which an extremum of~$f$ with respect to~$h$
crosses the zero line.  Denoting this point by $(h,z)=(a,b)$ and performing
an expansion about it to leading order in both~$h$ and~$z$, we then have
\begin{equation}
f(h,z) = {\partial f\over\partial z} (z-b) + {\partial f\over\partial h^2}
(h-a)^2 + \ldots,
\end{equation}
the terms in $f(a,b)$ and $\partial f/\partial h$ vanishing at the
extremum.  In the limit as we approach the band edge, therefore, the
equation~$f(h,z)=0$ takes the form
\begin{equation}
{\partial f\over\partial z} (z-b) + {\partial f\over\partial h^2} (h-a)^2 =
0,
\label{eq:fquadratic}
\end{equation}
and hence, within the band, we have $h(z) = a + \ii B\sqrt{b-z}$ for some
real constant~$B$.  Then the spectral density, Eq.~\eqref{eq:fptsolution2},
is
\begin{equation}
\rho(z) = C {\sqrt{b-z}\over z},
\end{equation}
where $C$ is another real constant.  A similar argument implies square-root
behavior at the lower edge of the spectrum as well.  The square-root form
can be seen, for example, in the vertical sides of the spectrum in
Fig.~\ref{fig:d2spect}.

We can also calculate the behavior of~$h(z)$ as $z\to b$ from above, for
which Eq.~\eqref{eq:fquadratic} implies
\begin{equation}
h(z) = a + B\sqrt{z-b},
\end{equation}
with the same real constant~$B$ as before.  Note that this implies that the
limiting value of $h(z)$ at the band edge is generically finite, but that
the slope $\dd h/\dd z$ diverges.  This has important consequences for
``hub'' vertices---those with unusually high degree---whose effect on the
spectrum displays a phase transition behavior that depends crucially on the
functional form of~$h(z)$.  We discuss hub vertices in detail in
Section~\ref{sec:hubs}.

These results apply for the case where the expected degree distribution is
bounded.  In cases where it is not we expected the spectral density of the
modularity matrix to be similarly unbounded, having no band edge and
generically inheriting the worst-case tail behavior of~$p(k)$.  Similar
observations have been made previously by Chung~\etal~\cite{CLV03} for a
different matrix, the graph Laplacian.  They note that a normalized version
of the Laplacian, akin to our normalized modularity matrix~$\widetilde{B}$,
should display a semicircle distribution, but that the Laplacian itself
should have a spectrum that inherits the tail behavior of the degree
distribution.

\section{The resolvent and the Stieltjes transform}
\label{sec:alt}
In the previous section we calculated the spectral density of the
modularity matrix for the configuration model.  It is possible to calculate
many other properties of the spectrum as well, as we now show.  Our
starting point for these calculations is the so-called resolvent matrix,
which is the matrix function~$\mat{R}(z) = (z\mat{I}-\mat{B})^{-1}$, where
$\mat{I}$ is the identity.  As we will see, a knowledge of the ensemble
average of the resolvent allows us to calculate many things, including the
spectral density of the adjacency matrix, the leading eigenvalue of the
adjacency matrix, and the effect on the spectrum of network hubs.

It also gives us an alternative, though perhaps less elegant, derivation of
the results for the modularity matrix in the previous section.  The
spectral density~$\rho(z)$ of the modularity matrix can be defined as
\begin{equation}
\rho(z) = {1\over n} \sum_{i=1}^n \delta(z-\lambda_i),
\label{eq:rhoz}
\end{equation}
where $\lambda_i$ are the eigenvalues of the matrix.  Substituting for the
delta-function from Eq.~\eqref{eq:delta}, we get the so-called
Plemelj--Sokhotski formula
\begin{equation}
\rho(z) = -{1\over n\pi} \lim_{\eta\to0+} \Im \sum_{i=1}^n
           {1\over z-\lambda_i+\ii\eta}.
\label{eq:gzinversion}
\end{equation}
Via a change of basis, the sum on the right-hand side is equal to the trace
of the matrix $[(z+\ii\eta)\mat{I}-\mat{B}]^{-1}$, and hence $\rho(z)$~is
the limit where $z$ goes to the real line of $-(1/n\pi)
\Im\,\tr(z\mat{I}-\mat{B})^{-1}$.  In other words, the spectral density
depends on the trace of the resolvent, and its average over the ensemble of
model networks is given by the average of this quantity:
\begin{equation}
\rho(z) = -{1\over n\pi}
  \Im \tr\bigl\langle(z\mat{I}-\mat{B})^{-1}\bigr\rangle.
\label{eq:density}
\end{equation}
The normalized trace $\tr(z\mat{I}-\mat{B})^{-1}/n$ is called the
\defn{Stieltjes transform} of~$\mat{B}$.

The two most common ways to calculate the Stieltjes transform are either to
expand the matrix $(z\mat{I}-\mat{B})^{-1}$ in powers of~$\mat{B}$ and take
the trace term by term, or to write the trace in terms of derivatives of a
Fresnel integral and then employ the replica trick~\cite{EJ76}.  Here,
however, we take a different approach inspired by work of Bai and
Silverstein~\cite{bai1999methodologies,bai2010spectral} that allows us to
calculate the average of the full resolvent.

The resolvent is the inverse of a matrix whose off-diagonal elements are
zero-mean random variables.  Consider a general such matrix~$\mat{X}$ and
let us write it in terms of its first $n-1$ rows and columns, plus the last
row and column, thus:
\begin{equation}
\mat{X} = \begin{pmatrix}
  \boxed{\begin{matrix} \\ \\ \qquad\mat{X}_n\qquad \\ \\ \\ \end{matrix}}
  &
  \boxed{\begin{matrix} \\ \phantom{m} \\ \vec{a} \\ \\ \null \end{matrix}}
  \\
  \boxed{\qquad\vec{a}^T\qquad} & x_{nn}\rule{0pt}{16pt} \\
          \end{pmatrix}
\end{equation}
Thus $\mat{X}_n$ is the matrix~$\mat{X}$ with the $n$th row and column
removed, and $\vec{a}$ is the $n$th column minus its last element~$x_{nn}$.

Now consider the vector $\vec{v} = \mat{X}^{-1}\vec{u}$, where
$\vec{u}=(0,\ldots,0,1)$.  Let us break $\vec{v}$ into its first $n-1$
elements and its last element~$\vec{v}=(\vec{v}_1|v_n)$, where clearly $v_n
= \bigl[ \mat{X}^{-1} \bigr]_{nn}$.  Then we have $\mat{X}\vec{v} =
\vec{u}$ and hence
\begin{equation}
\mat{X}_n \vec{v}_1 + v_n \vec{a} = 0,\qquad
\vec{a}^T\vec{v}_1 + x_{nn} v_n = 1.
\end{equation}
The first equation tells us that
\begin{equation}
\vec{v}_1 = - v_n \mat{X}_n^{-1}\vec{a},
\label{eq:offdiag}
\end{equation}
and substituting this result into the second gives
\begin{equation}
\bigl[ \mat{X}^{-1} \bigr]_{nn} = v_n
  = {1\over x_{nn} - \vec{a}^T\mat{X}_n^{-1}\vec{a}}.
\label{eq:tao1}
\end{equation}

To make further progress we assume that $v_n$ is narrowly peaked about its
average value in the limit of large system size, meaning its variance about
that value vanishes as $n$ becomes large.  We will for the moment take this
assumption as given, but it can be justified using results for
concentration of measure of random quadratic
forms~\cite{karoui2011geometric}, which apply provided vertex degrees are
large (so that our results, like those of Section~\ref{sec:spectrum}, will
be exact only for large degrees).

If $v_n$ is narrowly peaked then the average of the reciprocal on the
right-hand side of~\eqref{eq:tao1} is equal to the reciprocal of the
average and
\begin{equation}
\bigl\langle \bigl[ \mat{X}^{-1} \bigr]_{nn} \bigr\rangle
  = {1\over\av{x_{nn}} - \av{\vec{a}^T\mat{X}_n^{-1}\vec{a}}},
\label{eq:tao2}
\end{equation}
Furthermore, if~$v_n$ is narrowly peaked then the average of
Eq.~\eqref{eq:offdiag} is $\av{\vec{v}_1} = -v_n \av{\mat{X}_n}
\av{\vec{a}} = 0$ since $\vec{a}$ is independent of~$\mat{X}_n$ and
$\av{\vec{a}}=0$.  But the elements of $\vec{v}_1$ are equal
to~$[\mat{X}^{-1}]_{in}$ and hence
\begin{equation}
\bigl\langle \bigl[ \mat{X}^{-1} \bigr]_{in} \bigr\rangle = 0
\end{equation}
for $i\ne n$.  By the same method we can derive expressions for the inverse
of $\mat{X}$ with any row and column~$i$ removed and hence show that
\begin{equation}
\bigl\langle \bigl[ \mat{X}^{-1} \bigr]_{ii} \bigr\rangle
  = {1\over\av{x_{ii}} - \av{\vec{a}^T\mat{X}_i^{-1}\vec{a}}}
\label{eq:tao3}
\end{equation}
and
\begin{equation}
\bigl\langle \bigl[ \mat{X}^{-1} \bigr]_{ij} \bigr\rangle = 0\qquad
\mbox{for $i\ne j$.}
\label{eq:xdiag}
\end{equation}
In other words, $\av{\mat{X}^{-1}}$ is a diagonal matrix when $n$ is large,
with diagonal elements given by Eq.~\eqref{eq:tao3}.

But if this is true of $\mat{X}^{-1}$, then by the same argument it must
also be true of $\mat{X}_i^{-1}$.  Hence, noting that $\vec{a}$ is
independent of~$\mat{X}_i$, we have
\begin{equation}
\av{\vec{a}^T\mat{X}_i^{-1}\vec{a}}
  = \sum_{jk} \bigl\langle \bigl[ \mat{X}_i^{-1} \bigr]_{jk} \bigr\rangle
              \av{a_j a_k}
  = \sum_j \bigl\langle \bigl[ \mat{X}_i^{-1} \bigr]_{jj} \bigr\rangle
              \av{a_j^2}.
\label{eq:axa}
\end{equation}

Returning now to Eq.~\eqref{eq:density}, the role of the matrix~$\mat{X}$
in our problem is played by $z\mat{I}-\mat{B}$.  As we noted earlier, the
elements of the modularity matrix~$\mat{B}$ (and hence also the elements of
the vector~$\vec{a}$) have mean zero and variance~$k_ik_j/2m$.  Hence
$\av{a_j^2} = k_ik_j/2m$ in Eq.~\eqref{eq:axa} and
\begin{align}
\av{\vec{a}^T\mat{X}_i^{-1}\vec{a}} &=
   \sum_j \bigl\langle \bigl[ (z\mat{I}-\mat{B}_i)^{-1} \bigr]_{jj}
   \bigr\rangle {k_ik_j\over2m} \nonumber\\
  &= {k_i\over2m} \tr[\mat{D}_i\av{(z\mat{I}-\mat{B}_i)^{-1}}],
\label{eq:aziba}
\end{align}
where $\mat{D}$ is the diagonal matrix with elements~$k_i$ and $\mat{D}_i$
is the same matrix with the $i$th row and column removed.  

However, if $\tr\bigl[\mat{D}_i\av{(z\mat{I}-\mat{B}_i)^{-1}}\bigr]/2m$
tends to a well-defined limit as the network becomes large, then in this
limit it must equal
$\tr\bigl[\mat{D}\av{(z\mat{I}-\mat{B})^{-1}}\bigr]/2m$---the omission, or
not, of the $i$th row and column makes a vanishing difference for
large~$n$.  Hence~\eqref{eq:tao3} becomes
\begin{equation}
\bigl\langle \bigl[ (z\mat{I}-\mat{B})^{-1} \bigr]_{ii} \bigr\rangle
  = {1\over z - k_i \tr[\mat{D}\av{(z\mat{I}-\mat{B})^{-1}}]/2m},
\label{eq:basic2}
\end{equation}
where we have made use of the fact that $\av{B_{ii}} = 0$.  At the same
time, the off-diagonal elements of $\av{(z\mat{I}-\mat{B})^{-1}}$ are zero
by Eq.~\eqref{eq:xdiag}, so the average of the resolvent matrix is
diagonal, a result that will be crucial for several following developments.

Without loss of generality, we now label the vertices of our network in
order of increasing expected degree, and for convenience we define
functions~$\gamma_z(x)$ and $k(x)$ of the continuous variable~$x$ thus:
\begin{equation}
\gamma_z(i/n) = \bigl\langle\bigl[ (z\mat{I}-\mat{B})^{-1} \bigr]_{ii}
           \bigr\rangle,
\qquad
k(i/n) = k_i.
\label{eq:defsgamma}
\end{equation}
Then for large~$n$ Eq.~\eqref{eq:basic2} becomes
\begin{equation}
\gamma_z(x) = {1\over z - [k(x)/c]
                            \int_0^1 k(y) \gamma_z(y) \>\dd y},
\label{eq:qzx}
\end{equation}
where $c=2m/n$ is the average degree, as previously.

The spectral density, Eq.~\eqref{eq:density}, is related to $\gamma_z(x)$
by
\begin{equation}
\rho(z) = - {1\over\pi} \Im g(z),
\label{eq:rhogz}
\end{equation}
where
\begin{equation}
g(z) = {1\over n} \tr\bigl\langle(z\mat{I}-\mat{B})^{-1}\bigr\rangle
     = \int_0^1 \gamma_z(x) \>\dd x,
\label{eq:gz}
\end{equation}
which is just the ensemble average of the Stieltjes transform.  To
calculate~$g(z)$, we define the additional quantity
\begin{align}
h(z) &= {1\over2m}
        \tr \bigl[ \mat{D} \bigl\langle (z\mat{I}-\mat{B})^{-1}
            \bigr\rangle \bigr]
      = {1\over c} \int_0^1 k(x) \gamma_z(x) \>\dd x \nonumber\\
     &= {1\over c} \int_0^1 {k(x)\>\dd x\over z - k(x)h(z)},
\label{eq:solnh1}
\end{align}
where we have used Eq.~\eqref{eq:qzx}.  Since we have labeled our vertices
in order of increasing degree, $k(x)$~is by definition the $(nx)$th-lowest
degree in the network, or equivalently it is the functional inverse of the
cumulative distribution function~$P(k)$ defined by
\begin{equation}
P(k) = \int_0^k p(k') \>\dd k',
\end{equation}
where $p(k)$ is the expected degree distribution.  Thus, changing variables
from~$x$ to~$k$, Eq.~\eqref{eq:solnh1} can be written
\begin{equation}
h(z) = {1\over c} \int_0^\infty {k\>\dd P(k)\over z - k h(z)},
\end{equation}
or as either of the equivalent forms
\begin{equation}
h(z) = {1\over c} \int_0^\infty {k\,p(k)\>\dd k\over z - k h(z)}
     = \int_0^\infty {q(k)\>\dd k\over z - k h(z)},
\label{eq:solnh2}
\end{equation}
where the (correctly normalized) probability distribution
\begin{equation}
q(k) = {k\,p(k)\over c}
\label{eq:excess}
\end{equation}
is known as the \defn{excess degree distribution} in the networks
literature.  This distribution, which arises often in the theory of
networks, is the probability that the network vertex reached by following
an edge has an expected number $k$ of edges attached to it other than the
one we followed to reach the vertex.  (The distribution looks slightly
different from the form usually given~\cite{NSW01} because it is expressed
in terms of expected degree rather than actual degree.)

If we can solve Eq.~\eqref{eq:solnh2} for~$h(z)$ then we can calculate
$g(z)$ by substituting Eq.~\eqref{eq:qzx} into Eq.~\eqref{eq:gz} and again
changing variables from~$x$ to~$k$, to get
\begin{equation}
g(z) = \int_0^\infty {p(k)\>\dd k\over z - k h(z)}.
\end{equation}
This equation is similar in form to Eq.~\eqref{eq:solnh2}, but note that it
is the ordinary degree distribution~$p(k)$ that appears in the numerator,
not the excess degree distribution.

Alternatively, and more directly, we can calculate $g(z)$ by multiplying
both sides of~\eqref{eq:qzx} by the right-hand denominator, integrating,
and rearranging, to get
\begin{equation}
g(z) = {1 + c h^2(z)\over z}.
\label{eq:gzhz}
\end{equation}
Combining this result with Eq.~\eqref{eq:rhogz} now gives us the spectral
density:
\begin{equation}
\rho(z) = - {c\over\pi z} \Im h^2(z),
\label{eq:rhohz}
\end{equation}
where the imaginary part is, if necessary, calculated as the limit where
$z$ tends to the real line from above.

Equations~\eqref{eq:solnh2} and~\eqref{eq:rhohz} are precisely the
equations, \eqref{eq:fptsolution1}~and~\eqref{eq:fptsolution2}, that we
derived previously using the free convolution.

\section{Spectrum of the adjacency matrix}
In the previous sections we have derived the spectral density of the
modularity matrix.  To calculate the corresponding quantity for the
adjacency matrix we make use of an argument of~\cite{CDF09,BN11} as
follows.  The adjacency matrix can be written in terms of the modularity
matrix as $\mat{A} = \mat{B} + \vec{k}\vec{k}^T/2m$, where $\vec{k}$ is the
$n$-element vector with elements~$k_i$.  Hence any eigenvalue/vector pair
$z,\vec{v}$ of the adjacency matrix satisfies
\begin{equation}
\biggl( \mat{B} + {\vec{k}\vec{k}^T\over2m} \biggr) \vec{v}
  = z\vec{v},
\label{eq:theta1}
\end{equation}
which can be rearranged to read
\begin{equation}
{\vec{k}^T\vec{v}\over2m} (z\mat{I}-\mat{B})^{-1}\vec{k} = \vec{v}.
\end{equation}
Multiplying by~$\vec{k}^T$, we then find that
\begin{equation}
{1\over2m} \vec{k}^T (z\mat{I}-\mat{B})^{-1} \vec{k} = 1.
\label{eq:bounds1}
\end{equation}
Expanding~$\vec{k}$ as a linear combination of the eigenvectors~$\vec{b}_i$
of~$\mat{B}$, this result can be written
\begin{equation}
{1\over2m} \sum_{i=1}^n {(\vec{k}^T\vec{b}_i)^2\over z-\beta_i} = 1,
\label{eq:bounds2}
\end{equation}
where $\beta_i$ are the eigenvalues of the modularity matrix.

\begin{figure}
\begin{center}
\includegraphics[width=\columnwidth]{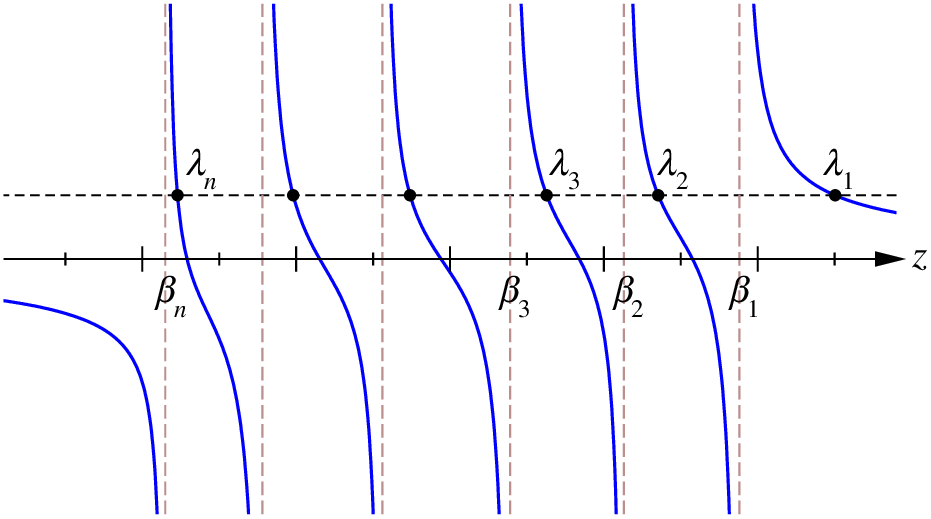}
\end{center}
\caption{The solutions~$\lambda_i$ to Eq.~\eqref{eq:bounds2} correspond to
  the points where the left-hand side of the equation (solid curves)
  equals~1 (dashed horizontal line).  This implies that the values of the
  $\lambda_i$ are interleaved between the eigenvalues~$\beta_i$ of the
  modularity matrix.}
\label{fig:interleave}
\end{figure}

The solutions of this equation can be visualized as in
Fig.~\ref{fig:interleave}.  The solid curves represent the left-hand side
of the equation, which has poles as shown at $z=\beta_i$ for all~$i$.  The
dashed horizontal line represents the 1 on the right-hand side and the
points at which it intercepts the curves are the solutions for~$z$
of~\eqref{eq:bounds2}, which are the eigenvalues~$\lambda_i$ of the
adjacency matrix.  If we number the eigenvalues of both $\mat{A}$ and
$\mat{B}$ in order from largest to smallest, the geometry of
Fig.~\ref{fig:interleave} implies that the eigenvalues must satisfy an
interleaving condition of the form $\lambda_1 \ge \beta_1 \ge \lambda_2 \ge
\beta_2\ge \ldots \ge \lambda_n \ge \beta_n$.  In the limit of large~$n$,
where the spectral density of the modularity matrix becomes a smooth
function and the eigenvalues are arbitrarily closely spaced, this implies
that $\lambda_i\to\beta_i$, so that asymptotically the spectral density of
the adjacency matrix is the same as that of the modularity matrix.

The only exception is the highest eigenvalue of the adjacency
matrix~$\lambda_1$, which is bounded below by~$\beta_1$, but unbounded
above.  To calculate this value we average~\eqref{eq:bounds1} over the
ensemble and recall, as demonstrated in Section~\ref{sec:alt}, that
$\av{(z\mat{I}-\mat{B})^{-1}}$ is diagonal, and hence
\begin{equation}
{1\over2m} \vec{k}^T \bigl\langle (z\mat{I}-\mat{B})^{-1} \bigr\rangle
  \vec{k} = {1\over2m} \sum_i k_i^2 \bigl\langle \bigl[
             (z\mat{I}-\mat{B})^{-1} \bigr]_{ii} \bigr\rangle.
\label{eq:theta2}
\end{equation}
Combining this result with~\eqref{eq:bounds1} and using
Eq.~\eqref{eq:defsgamma} we then have
\begin{equation}
{1\over c} \int_0^1 k^2(x) \gamma_z(x) \>\dd x = 1.
\label{eq:leading1}
\end{equation}
Taking Eq.~\eqref{eq:qzx}, multiplying by the right-hand denominator and a
further factor of~$k(x)$, then integrating over~$x$, we get
\begin{equation}
c z h(z) - h(z) \int_0^1 k^2(x) \gamma_z(x) \>\dd x =
  \int_0^1 k(x) \>\dd x.
\end{equation}
And combining this result with Eq.~\eqref{eq:leading1} and noting that
$\int_0^1 k(x) \>\dd x = \int_0^\infty k\,p(k)\>\dd k = c$, we have
\begin{equation}
(z-1) h(z) = 1.
\label{eq:leading2}
\end{equation}
The solution of this equation for~$z$ gives us the leading
eigenvalue~$\lambda_1$ of the adjacency matrix.

For the Poisson random graph, for example, this result, in combination with
Eq.~\eqref{eq:poissonhz}, tells us that the leading eigenvalue takes the
value $c+1$.  This is not a new result---it is well known in the
literature---but it is comforting to see that the formalism works.

For the two-degree model of Eq.~\eqref{eq:twodegree}, we can
use~\eqref{eq:leading2} to eliminate~$h(z)$ from~\eqref{eq:twodegree} and
get
\begin{equation}
{p_1d_1+p_2d_2\over(z-1)^2} = {p_1d_1\over z(z-1)-d_1}
                              + {p_2d_2\over z(z-1)-d_2},
\label{eq:d2leading}
\end{equation}
which gives us a cubic equation for~$z$.  For the parameter values used in
Fig.~\ref{fig:d2spect}, for example, $d_1=50$, $d_2=100$, and
$p_1=1-p_2=\frac14$, we find that the leading eigenvalue of the adjacency
matrix is $z=93.893\ldots$\ \ A~numerical calculation for the same
parameters is in good agreement, giving $z=93.896\pm0.017$ for an average
over 100 systems of size $n=10\,000$.

For the case of general degree distribution, we can use~\eqref{eq:leading2}
to eliminate $h(z)$ in Eq.~\eqref{eq:solnh2} to get
\begin{equation}
{z\over z-1}
  = \int_0^\infty {q(k)\>\dd k\over 1 - k/(z^2-z)}.
\end{equation}
An exact solution to this equation requires us to perform the integral, but
one can derive an approximate solution by expanding the denominator of the
integrand:
\begin{equation}
{z\over z-1}
  = 1 + \int_0^\infty \sum_{r=1}^\infty {k^r\over (z^2-z)^r}\>q(k)\>\dd k,
\end{equation}
or
\begin{equation}
{1\over z-1}  = \sum_{r=1}^\infty {\av{k^r}_q\over(z^2-z)^r},
\end{equation}
where $\av{\ldots}_q$ denotes an average over the excess degree
distribution of Eq.~\eqref{eq:excess}.  If $z^2-z\gg k_\textrm{max}$, where
$k_\textrm{max}$ is the largest degree in the network, and noting that
$\av{k^r}_q\le k_\textrm{max}^r$, we have
\begin{equation}
{1\over z-1} = {\av{k}_q\over z^2-z}
  + \Ord\bigl[k_\textrm{max}/(z^2-z)\bigr]^2,
\end{equation}
or
\begin{equation}
z \simeq {\av{k^2}\over\av{k}}
\label{eq:clv}
\end{equation}
to leading order, where we have made use of $\av{k}_q=\av{k^2}/\av{k}$.
This result was derived previously by other means by
Chung~\etal~\cite{CLV03}.

Taking the example of the two degree model above again, this approximation
gives
\begin{equation}
z \simeq {p_1d_1^2+p_2d_2^2\over p_1d_1+p_2d_2},
\end{equation}
and for the parameter values of Fig.~\ref{fig:d2spect} we find that
$z\simeq 92.86$, which differs by about 1\% from the true value of $93.89$
given by Eq.~\eqref{eq:d2leading}.

\section{Network hubs}
\label{sec:hubs}
The picture developed in the previous sections is one in which the spectrum
of the adjacency matrix has two primary components: a single leading
eigenvalue plus a continuous band of lower eigenvalues, which it shares
with the modularity matrix.

Let us examine more closely the continuous band and concentrate on the case
of the modularity matrix, which is simpler since it has only the band and
no separate leading eigenvalue.  Consider the eigenvalues that lie at the
topmost edge of the band, which are the highest eigenvalues of the
modularity matrix.  These eigenvalues are normally associated with good
bisections of the network into ``communities''---if a good bisection exists
then there will be a corresponding high-lying eigenvalue whose
eigenvector's elements describe the split~\cite{Newman06b}.

As we now argue, however, there is another mechanism that generates
high-lying eigenvalues, namely the presence of hubs in the
network---vertices of unusually high degree---and the highest eigenvalues
in the spectrum of the modularity matrix, and also the lowest, are often
due to these hubs, while those corresponding to communities are somewhat
smaller.  As we will see, for hubs of sufficiently high degree, these
eigenvalues can split off from the continuous band in a manner reminiscent
of impurity states in condensed matter physics.  In effect, the hub acts as
an impurity in the network.

To see how the addition of a hub to a network produces a high-lying
eigenvalue, let the hub be vertex~$n$ and let $\mat{B}_n$ once again be the
modularity matrix without the~$n$th vertex (i.e.,~with the $n$th row and
column removed), so that the full modularity matrix looks like this:
\begin{equation}
\mat{B} = \begin{pmatrix}
  \boxed{\begin{matrix} \\ \\ \qquad\mat{B}_n\qquad \\ \\ \\ \end{matrix}}
  &
  \boxed{\begin{matrix} \\ \phantom{m} \\ \vec{a} \\ \\ \null \end{matrix}}
  \\
  \boxed{\qquad\vec{a}^T\qquad} & b_{nn}\rule{2pt}{0pt}\rule{0pt}{16pt} \\
          \end{pmatrix}.
\end{equation}

Now, in an argument analogous to that of the previous section, consider an
eigenvector of this matrix $\vec{v} = (\vec{v}_1| v_n)$.  Then the
eigenvector equation~$\mat{B}\vec{v} = z\vec{v}$ can be multiplied out to
give the equations
\begin{align}
\label{eq:hub1}
\mat{B}_n \vec{v}_1 + v_n \vec{a} &= z\vec{v}_1, \\
\label{eq:hub2}
\vec{a}^T\vec{v}_1 + b_{nn} v_n &= zv_n.
\end{align}
The first of these can be rearranged to give
\begin{equation}
\vec{v}_1 = v_n (z\mat{I}-\mat{B}_n)^{-1} \vec{a}.
\label{eq:v1}
\end{equation}
Then multiplying by $\vec{a}^T$ and using the second equation gives
\begin{equation}
\vec{a}^T (z\mat{I}-\mat{B}_n)^{-1} \vec{a} = z - b_{nn}.
\end{equation}
Now we note that the $i$th element of~$\vec{a}$ is an independent random
variable with variance $k_nk_i/2m$ and we can average over the ensemble and
apply Eq.~\eqref{eq:axa} to rewrite the left-hand side, giving
\begin{equation}
{k_n\over2m} \tr \bigl[ \mat{D}_n \bigl\langle 
                        (z\mat{I}-\mat{B}_n)^{-1} \bigr\rangle \bigr] = z,
\label{eq:hub3}
\end{equation}
where $\mat{D}$ is the diagonal matrix with elements~$k_i$ as before,
$\mat{D}_n$~is the same matrix with the~$n$th row and column removed, and
we have made use of the fact that $\av{b_{nn}}=0$.  We note, as previously,
that if the quantity $\tr (\mat{D}_n \av{(z\mat{I}-\mat{B}_n)^{-1}})/2m$
tends to a limit as the network becomes large, then that limit is equal to
the function~$h(z)$ defined in Eq.~\eqref{eq:solnh1}.  Thus the
eigenvalue~$z$ satisfies
\begin{equation}
h(z) = {z\over k_n}.
\label{eq:hub4}
\end{equation}
Substituting this expression into Eq.~\eqref{eq:solnh2} and rearranging, we
get an explicit expression for the eigenvalue thus:
\begin{equation}
z^2 = {k_n^2\over c} \int_0^\infty {k\,p(k)\>\dd k\over k_n-k}.
\label{eq:hubz}
\end{equation}
This calculation also extends to the case where there is more than one hub
in the network.  Because the hub is treated no differently from any other
network vertex, the same arguments apply if we add a second hub, or more,
after the first.  Equation~\eqref{eq:hubz} will give the correct eigenvalue
for each hub separately.

Once again, our ability to actually solve for the value of~$z$ will depend
on whether we can do the integral in Eq.~\eqref{eq:hubz} (although one
could also evaluate the integral numerically).  In the special case where
the hub degree~$k_n$ is much larger than the expected degree of any of the
other vertices, so that $k_n-k\simeq k_n$ in the denominator of the
integrand, the expression simplifies to
\begin{equation}
z^2 = {k_n\over c} \int_0^\infty k\,p(k)\>\dd k = k_n,
\label{eq:largekn}
\end{equation}
and hence $z = \sqrt{k_n}$.

\begin{figure}
\begin{center}
\includegraphics[width=8cm]{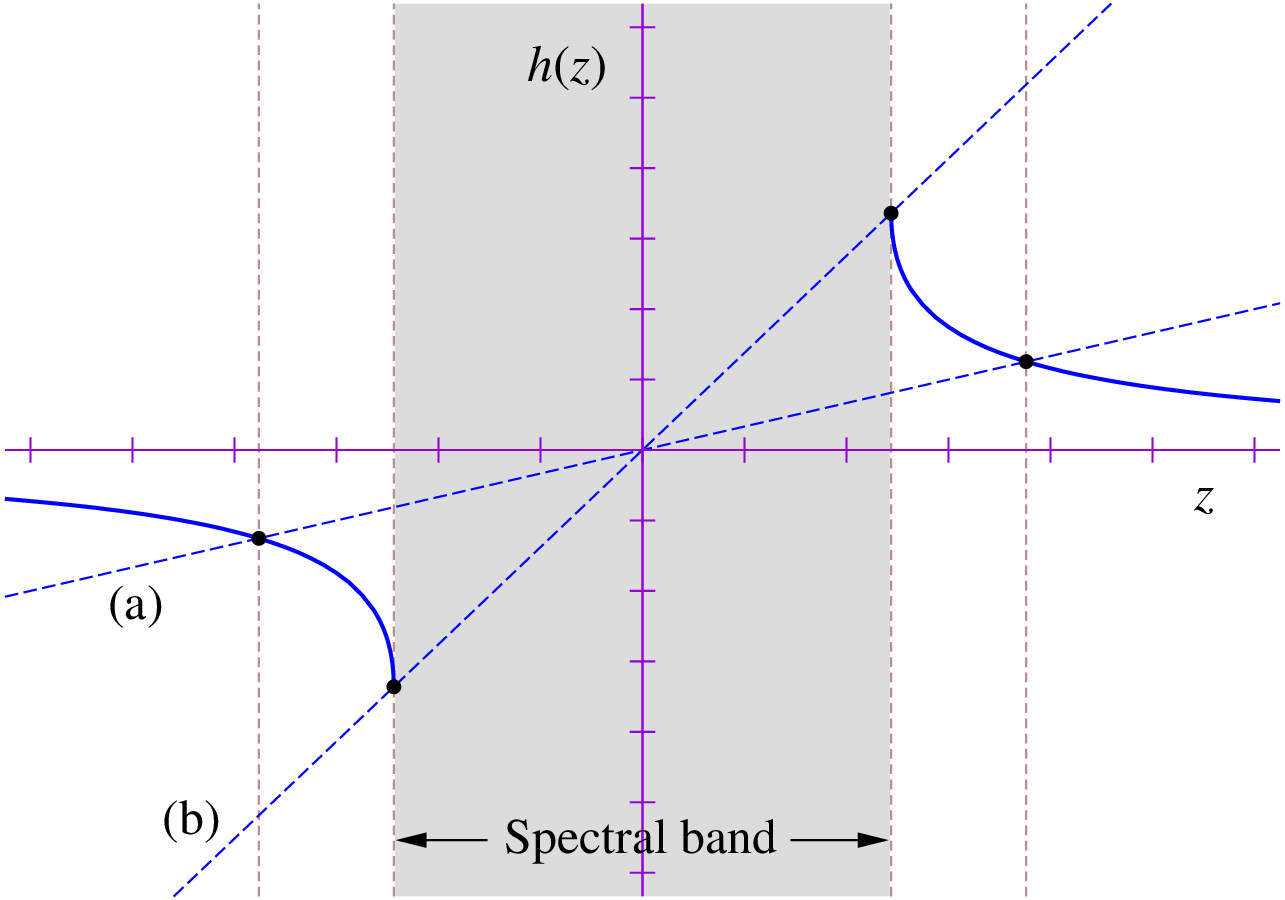}
\end{center}
\caption{Graphical solution of Eq.~\eqref{eq:hub4}.  The solid curves
  represent the value of $h(z)$ as a function of~$z$, above and below the
  spectral band, and the hub eigenvalues, which are solutions of
  Eq.~\eqref{eq:hub4}, fall at the points where this curve intersects the
  straight line~$z/k_n$, represented by the dashed diagonal line.  The
  slope of this line is $1/k_n$, and hence when $k_n$ is large enough the
  lines intersect---case~(a)---and we have two hub eigenvalues, one above
  and one below the band (marked by dots).  Case~(b) is the borderline
  case.  If $k_n$ is any less than this value then there is no intersection
  and the highest and lowest eigenvalues will be those at the band edges.}
\label{fig:hubcalc}
\end{figure}

The solutions of Eq.~\eqref{eq:hub4} can be represented graphically as in
Fig.~\ref{fig:hubcalc}.  The curves in the figure represent the
function~$h(z)$ and the diagonal lines represent $z/k_n$.  The point where
the two cross give the eigenvalues.  As the figure shows, when the expected
degree $k_n$ of the $n$th vertex is large enough, the equation has two
solutions, one for low~$z$ and one for high and both given by
Eq.~\eqref{eq:hubz}, that are separate from the continuous spectrum of
eigenvalues we calculated in Section~\ref{sec:spectrum}.

How high a degree does a hub have to have to generate eigenvalues of this
kind?  The answer can be seen from Fig.~\ref{fig:hubcalc}---$k_n$~must be
large enough for the line $z/k_n$ to intercept the curve of~$h(z)$.  Thus
there is a critical value of~$k_n$, represented by the steeper diagonal in
the figure, below which the hub eigenvalues vanish.  Below this point, the
highest eigenvalue will fall at the edge of the continuous band as normal
and there will be no special hub eigenvalue.  We can derive an expression
for the transition point by observing that, as shown in
Section~\ref{sec:features}, the slope of~$h(z)$ diverges at the band edge,
which implies that $\dd z/\dd k_n=0$.  Differentiating Eq.~\eqref{eq:hubz}
and setting the result to zero, we find that the critical value of~$k_n$ is
the solution of
\begin{equation}
\int_0^\infty {k\,p(k)\>\dd k\over k_n-k}
  = \int_0^\infty {k^2\,p(k)\>\dd k\over(k_n-k)^2}.
\label{eq:bandedge}
\end{equation}
For example, in the case of the Poisson random graph this implies that the
transition takes place at the point where $c/(k_n-c) = c^2/(k_n-c)^2$,
i.e.,~when $k_n=2c$.  Thus we must have $k_n > 2c$ for the hub to have an
effect on the spectrum.

This gives us a working definition of what we mean by a ``hub'' in a
network.  It depends, not surprisingly, on the degree distribution of the
rest of the network---what it takes to stand out in a crowd depends on the
rest of the crowd.  But in the Poisson random graph, for instance, a hub is
a hub, in spectral terms, if its degree is greater than twice the average
in the rest of the network.  This is a somewhat surprising result, given
that vertices of high degree are easily spotted long before this point is
reached, at least for large~$c$.  Since the standard deviation of the
degree distribution is~$\sqrt{c}$, a vertex with degree twice the mean is
$\sqrt{c}$ standard deviations above the mean, which is a large number for
large~$c$.

\begin{figure}
\begin{center}
\includegraphics[width=\columnwidth]{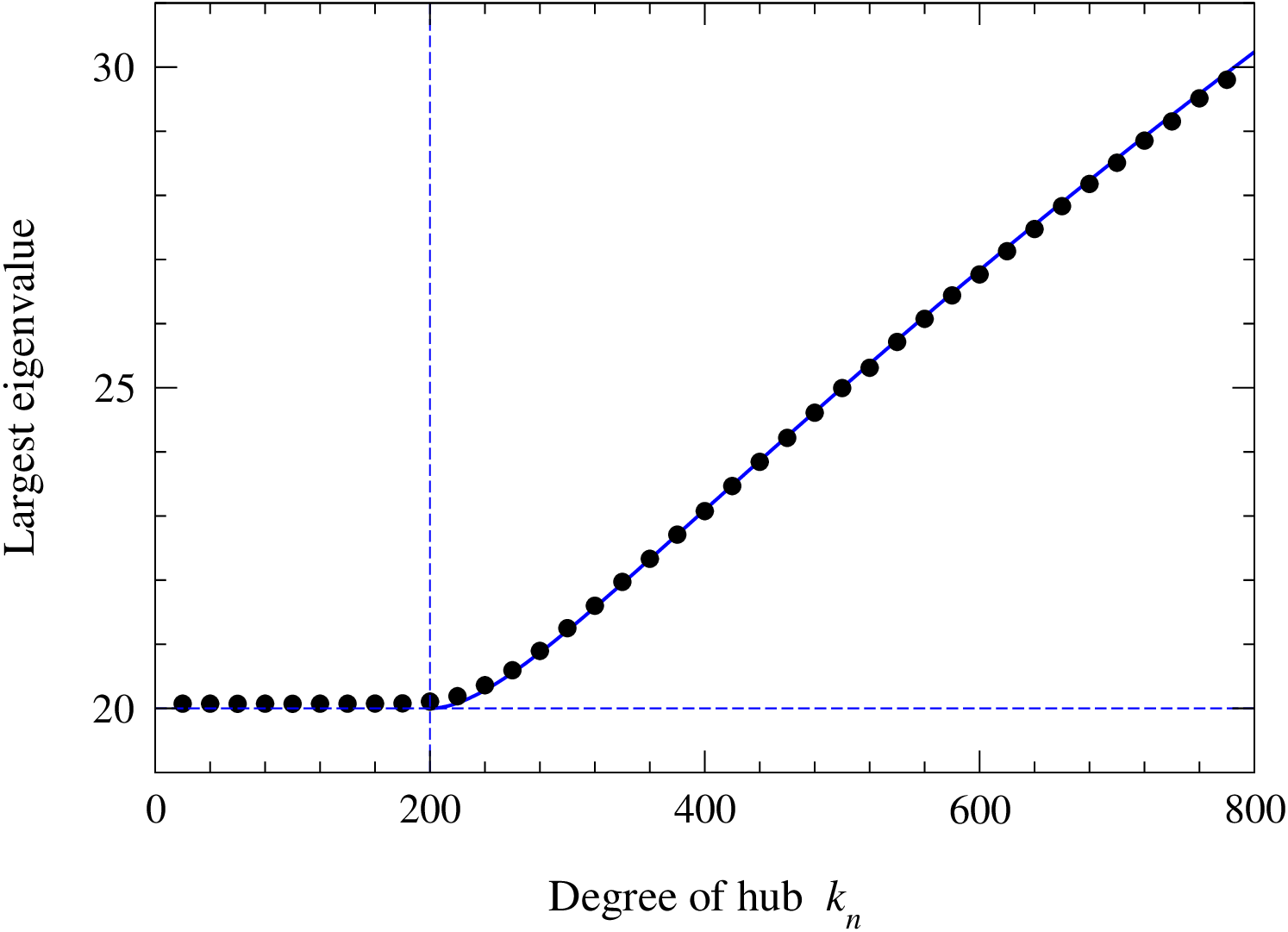}
\end{center}
\caption{The largest eigenvalue of the modularity matrix for a Poisson
  random graph of mean degree $c=100$ plus a single additional hub of
  expected degree~$k_n$.  Points are numerical results, averaged over 1000
  networks of $n=10\,000$ vertices each.  Statistical errors on the
  measurements are smaller than the points in all cases.  The solid curve
  is Eq.~\eqref{eq:hubz}, which gives $z=k_n/\sqrt{k_n-c}$ in this case,
  and the horizontal dashed line represents the value $z=2\sqrt{c}=20$,
  which is the lower limit on the eigenvalue set by the edge of the
  continuous spectral band.  The vertical dashed line represents the
  critical value $k_n=200$ of the hub degree, set by
  Eq.~\eqref{eq:bandedge}.}
\label{fig:hubresults}
\end{figure}

Nonetheless, the result does appear to be correct.
Figure~\ref{fig:hubresults} shows the results of numerical calculations of
the largest eigenvalue of the modularity matrix for a Poisson random graph
with a single additional hub of expected degree~$k_n$, as a function
of~$k_n$.  As the figure shows, the eigenvalue obeys Eq.~\eqref{eq:hubz}
quite closely until $k_n$ falls below~$2c$ (the vertical dashed line).
Past this point, the leading eigenvalue assumes the same value~$2\sqrt{c}$
as in a standard Poisson random graph with no hub (the horizontal line),
even though the hub may still be present.

Putting together our principal observations, we have now developed quite a
complete picture of the spectrum of the configuration model.  We expect the
spectrum to have two main parts, plus a third when the degree distribution
implies the presence of hubs:
\begin{enumerate}
\item There is a single eigenvalue given by the solution of
  Eq.~\eqref{eq:leading2}, which will normally be the leading eigenvalue.
\item There is a continuous band, given by Eq.~\eqref{eq:fptsolution2}.
  For bounded degree distributions the band will also be bounded, both
  above and below, and have edges that decay to zero as a square root.
\item If there are hubs in the network, then there will be additional
  eigenvalues outside the band at both ends, given by Eq.~\eqref{eq:hubz}.
  Each hub contributes two eigenvalues, one at each end of the band.
\end{enumerate}

\subsection{Localization around hubs}
One can also look at the eigenvector corresponding to a hub eigenvalue,
which turns out to be heavily localized around the hub vertex.  All the
elements of the eigenvector, except for the element~$v_n$ corresponding to
the hub itself, are given in terms of~$v_n$ by Eq.~\eqref{eq:v1}.  For
given~$\vec{a}$, the expected value of the $i$th component is
\begin{align}
v_i = v_n
   \Bigl[ \bigl\langle(z\mat{I}-\mat{B}_n)^{-1} \bigr\rangle \vec{a}
   \Bigr]_i
     = v_n \Bigl[ \bigl\langle(z\mat{I}-\mat{B}_n)^{-1} \bigr\rangle
   \Bigr]_{ii} a_i,
\end{align}
where we have once again made use of the fact that
$\av{(z\mat{I}-\mat{B}_n)^{-1}}$ is diagonal (see Eq.~\eqref{eq:xdiag}).

The $i$th element of the vector~$\vec{a}$ takes the value $a_i =
1-k_ik_n/2m$ for vertices~$i$ that are connected to the hub and
$-k_ik_n/2m$ for those that are not.  Hence, in the limit of large~$n$,
eigenvector elements corresponding to neighbors of the hub will be of order
a constant, with expected value
\begin{equation}
v_i = v_n \gamma_z(i/n) = {v_n\over z - k_i h(z)}
    = {v_n\over z(1-k_i/k_n)},
\label{eq:vi}
\end{equation}
with $z$ given by Eq.~\eqref{eq:hubz}, while the remaining elements will be
of order~$1/n$.

The value of~$v_n$ can be determined by insisting that the complete
eigenvector be normalized.  Using Eq.~\eqref{eq:v1} we can write the
normalization condition in the form
\begin{align}
1 &= |\vec{v}|^2 = v_n^2 + |\vec{v}_1|^2
   = v_n^2 \bigl[ 1 + \vec{a}^T (z\mat{I}-\mat{B})^{-2} \vec{a} \bigr]
     \nonumber\\
  &= v_n^2 \biggl[ 1 - {\dd\over\dd z} \vec{a}^T (z\mat{I}-\mat{B})^{-1}
                  \vec{a} \biggr].
\label{eq:vn1}
\end{align}
When we average over the ensemble we have, by analogy with
Eq.~\eqref{eq:aziba},
\begin{equation}
\bigl\langle \vec{a}^T (z\mat{I}-\mat{B})^{-1} \vec{a} \bigr\rangle =
{k_n\over2m} \tr \bigl\langle \mat{D}(z\mat{I}-\mat{B})^{-1} \bigr\rangle
  = k_n h(z),
\end{equation}
and hence~\eqref{eq:vn1} implies that
\begin{equation}
v_n^2 = {1\over 1 - k_n h'(z)},
\end{equation}
where $h'(z)$ denotes the first derivative of~$h(z)$, and we are assuming
once again that the vector element~$v_n$ is narrowly peaked about its
expected value.  Note that $h'(z)$ is negative at both the positive and
negative band edges, and diverges to $-\infty$ as we approach the band
edge.  Thus $v_n\to0$ as we approach the transition at the which the hub
eigenvalue disappears.

The results above apply to the hub eigenvectors at both ends of the
spectral band, there being two eigenvalues for each hub vertex, one at
either end, as shown in the previous section.  Both eigenvectors will have
a single element of order~1 in the position corresponding to the hub
itself, elements of order~$1/z$ in the positions corresponding the
neighbors of the hub (see Eq.~\eqref{eq:vi}), and all other elements of
order~$1/n$.  In other words, the both eigenvectors are strongly localized
in the neighborhood of the hub.  The only qualitative difference between
the two eigenvectors is in the sign of the elements corresponding to the
neighbors which, because of Eq.~\eqref{eq:vi}, will have the same sign
as~$v_n$ for the positive eigenvalue and the opposite sign for the negative
one.

\begin{figure}
\begin{center}
\includegraphics[width=\columnwidth]{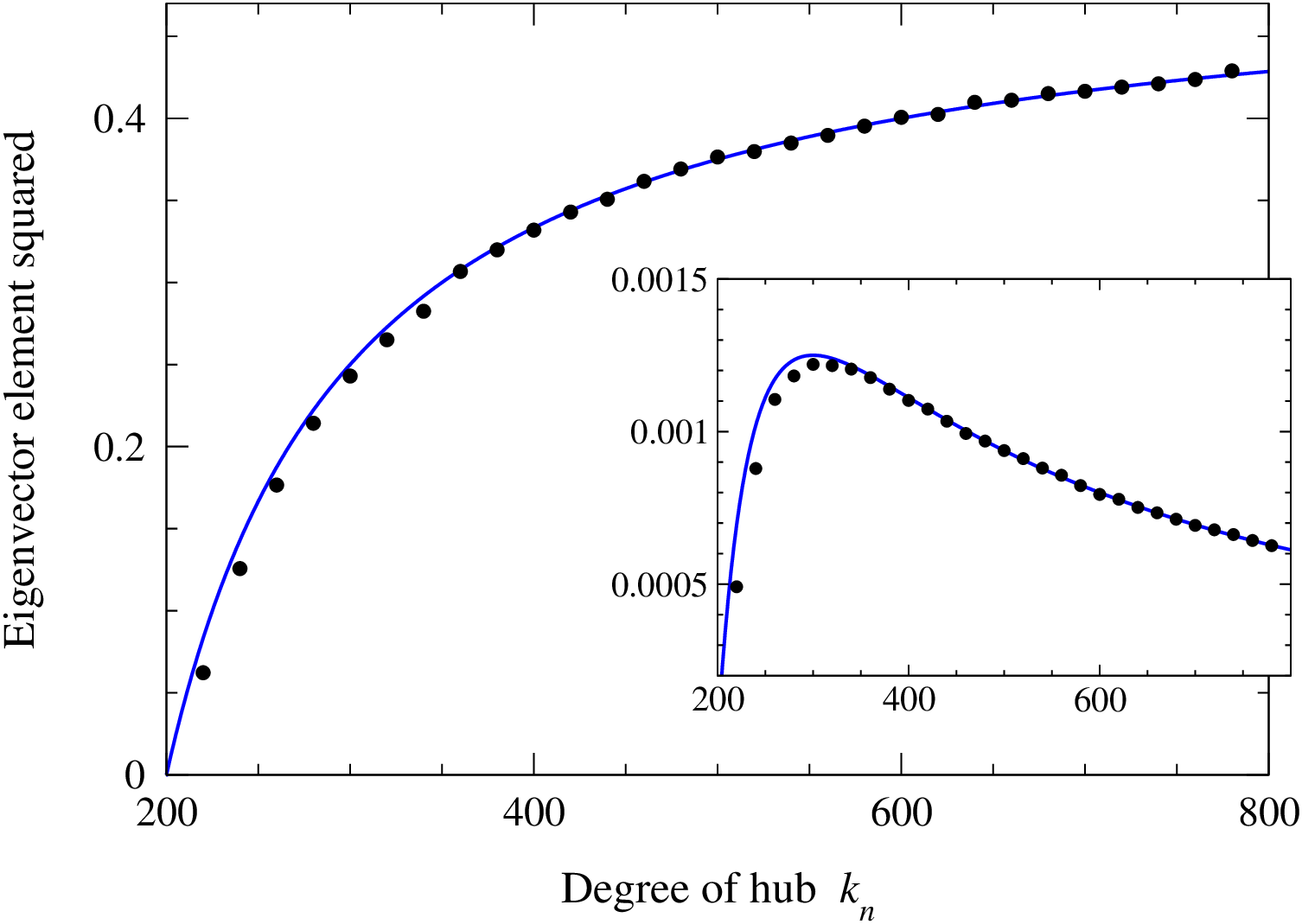}
\end{center}
\caption{Values of elements of the leading eigenvector of the modularity
  matrix for a Poisson random graph with $n=10\,000$ vertices and mean
  degree $c=100$, with a single added hub of degree~$k_n$.  Main figure:
  value of the vector element corresponding to the hub itself.  Inset:
  average value of the elements corresponding to the hub's immediate
  network neighbors.  Points are numerical results, averaged over 100
  networks each; curves are the analytic prediction, Eq.~\eqref{eq:hubev}.
  Statistical errors are smaller than the data points in all cases.}
\label{fig:hubev}
\end{figure}

As an example, consider again the Poisson random graph, for which $h(z)$ is
given by Eq.~\eqref{eq:poissonhz} and $z$ is given by Eq.~\eqref{eq:hubz}
to be $\pm k_n/\sqrt{k_n-c}$, so that $h'(z) = -1/(k_n-2c)$ and the
expected values of the eigenvector elements at both ends of the spectrum
satisfy
\begin{equation}
v_i^2 = \left\lbrace\begin{array}{ll}
          \bigl(\half k_n-c\bigr)/\bigl(k_n-c\bigr)
            & \qquad\mbox{for $i=n$,} \\[2pt]
          \bigl(\half k_n-c\bigr)/\bigl(k_n-c\bigr)^2
            & \qquad\mbox{for $i$ a neighbor
            of~$n$,} \\[4pt]
          0 & \qquad\mbox{otherwise,}
        \end{array}\right.
\label{eq:hubev}
\end{equation}
in the limit of large network size.  Figure~\ref{fig:hubev} shows a
comparison of these predictions with numerical results for actual networks.
As the figure shows, the agreement is once again good, although, as with
some of the other calculations, there are small disparities close to the
transition at which the hub eigenvalue meets the band edge (which is at
$k_n=200$ in this case).

\section{Conclusions}
In this paper we have studied the spectra of the adjacency and modularity
matrices of random networks with given expected degrees.  Our principal
findings are that the spectral densities of the adjacency and modularity
matrices are the same in the limit of large system size, except that the
adjacency matrix has an additional highest eigenvalue, and that the
spectral densities are given by the free multiplicative convolution of the
degree distribution with a Wigner semicircle distribution.  We have
confirmed these results with numerical studies of actual networks generated
according to the model.  The spectra show strong departures from the
classical semicircle law, in agreement with numerical studies by previous
authors.

We have also studied the effect of network hubs, vertices of unusually high
degree, and find that when their degree is sufficiently large these give
rise to eigenvalues outside the main band of the spectrum, akin to impurity
states in condensed matter systems.  We have derived an explicit formula
for these hub eigenvalues and we show that the corresponding eigenvectors
are strongly localized around the hubs themselves.

In addition to their relevance to partitioning, community structure, and
dynamical systems on networks, the techniques developed here could form a
starting point for spectral calculations in more elaborate networks.  There
has, for instance, been recent interest in the spectral properties of
community structured networks~\cite{CGO09,NN12}, but calculations have been
limited to models with Poisson degree distribution.  Applications of the
methods presented here to such networks could lead to new results for
structured networks with nontrivial degree distributions.

\begin{acknowledgments}
  The authors thank Lenka Zdeborova for useful conversations.  This work
  was funded in part by the National Science Foundation under grants
  CCF--1116115 and DMS--1107796 and by the Army Research Office under MURI
  grant W911NF--11--1--0391.
\end{acknowledgments}

\end{document}